\documentclass{sn-jnl}

\usepackage{graphicx}%
\usepackage{multirow}%
\usepackage{amsmath,amssymb,amsfonts}%
\usepackage{amsthm}%
\usepackage{mathrsfs}%
\usepackage[title]{appendix}%
\usepackage{xcolor}%
\usepackage{textcomp}%
\usepackage{manyfoot}%
\usepackage{booktabs}%
\usepackage{algorithm}%
\usepackage{algorithmicx}%
\usepackage{algpseudocode}%
\usepackage{listings}
\usepackage{comment}
\usepackage{lipsum}
\usepackage{makecell}
\usepackage{nicematrix}
\usepackage{slashed}
\usepackage{hyperref}
\usepackage{amsbsy}
\usepackage{bm}
\usepackage{changepage}
\usepackage[T2A,T1]{fontenc}
\usepackage[utf8]{inputenc}
\usepackage[russian,UKenglish]{babel}
\counterwithin*{equation}{section}
\usepackage[
backend=biber,
bibstyle=apa,
citestyle=numeric-comp,
sorting=none
]{biblatex}
\renewcommand\theequation{\thesection.\arabic{equation}}
\renewcommand\thefigure{\Alph{figure}}

\theoremstyle{thmstyleone}%
\theoremstyle{thmstyletwo}%

\theoremstyle{thmstylethree}%

\begin{document}

\title[Article Title]{Breaching the light barrier without paradoxes}


\author*[1]{\fnm{Abhishek} \sur{Mehta}}\email{abhishek.mehta@apctp.org, abhishek.mehta@khu.ac.kr}



\affil*[1]{
\orgname{APCTP}, \orgaddress{\street{77 Cheongam-ro, Nam-gu}, \city{Pohang-si}, \postcode{37673}, \state{Gyeongsangbuk-do}, \country{Korea}}}




\abstract{In this paper, we propose the regularization principle that resolves the temporal paradoxes associated with faster-than-light particles or tachyons at the macroscopic scale. The principle involves using the properties of the $\zeta$-function to regularize the unphysical momenta of tachyons that moves backwards in-time as an infinite sum of physical tachyon momenta that travel forwards in-time. Since the tachyon moves forward in-time in all Lorentz frames, this ensures that the tachyon processes are paradox-free without invoking the Reinterpretation Principle (RIP). Apart from resolving these paradoxes associated with faster-than-light particles, its other notable consequences especially pertaining to the censorship of naked singularities and faster-than-light communication are also discussed.}

\keywords{Tachyons, Black holes, Cosmic Censorship, Zeta function}



\maketitle

\section{Introduction}\label{sec1}
Tachyons refer to hypothetical particles that have coordinate speed exceeding the speed of light \cite{bilaniuk1969particles}. 
Due to the temporal inconsistencies associated with tachyons at the macroscopic scale \cite{Shoshany:2019yoz}, they are mostly considered unphysical and are therefore, eliminated from theories via intricate formalisms like the Higgs' mechanism \cite{peskin2018introduction} and supersymmetry \cite{Schwarz:2012zc} or by interpreting them as artifacts of an unstable or incorrect vacuum in a field theory \cite{Tong:2009np}. 
Despite that, several attempts are continually made to develop a consistent tachyonic mechanics \cite{bilaniuk1969particles, maccarrone1980, caldirola1980causality, recami1984classical}. The tachyon mechanics that are formulated exclusively rely on a frame-dependent reinterpretation of the tachyon processes where the emission of tachyons that move backwards in-time are reinterpreted as an absorption or vice-versa \cite{bilaniuk1969particles}. However, this reinterpretation is not sufficient to rule out the temporal paradoxes completely for more complicated tachyon processes \cite{Benford:1970xv}. It is also argued that the reinterpretation principle implies that the tachyon is more akin to a force-carrier and therefore, would only play a role as ``internal lines'' in particle interactions. This means that these paradoxes are no longer of any relevance as they assume the existence of free, modulated tachyon radiation to work \cite{recami1984classical, recami1985tachyons}. This understanding is also echoed in recent investigations where tachyons are incorporated in models of cosmology, dark matter and quantum gravity \cite{schwartz2025equation, mandal2025interacting, Ilkhchi:2025gbd}. Therefore, the reinterpretation principle inevitably leads to tachyons being restricted to quantum regimes with no classical analogues \cite{PhysRevD.110.015006}.\\

In this paper, we present a novel departure from the existing means of preserving tachyon causality in the literature. In order to describe a macroscopic tachyon mechanics, we instead propose a `regularization' principle of tachyons that move backwards in-time. We use momentum conservation to first identify emission and absorption processes involving only tachyons that move forwards in-time. This imposes a selection rule on the sign of the tachyon mass parameters depending on the process. The selection rule forces us to work with negative mass particles to ensure that the tachyon signals travel forwards in-time in the reference frames of emitters and receivers. To further avoid the RIP, we regularize the momenta of tachyons moving backwards in-time as an infinite sum of tachyon momenta that travel forward in-time which, referred to as the `tachyon shower', using the properties of the $\zeta$-function. This ensures that tachyons travel forwards in-time in all frames of references related by Lorentz transformations either as a single particle or a shower of particles effectively resolving all temporal paradoxes. This has an interesting consequence on faster-than-light communications i.e. in certain frames information encoded in tachyon signals may be become unintelligible due to Inter-Symbol Interference (ISI). Additionally, we demonstrate that paradox-free tachyons play an essential role to censorship of naked singularities in General Relativity (GR).\\

\section{Two-way Tachyonic antitelephone}
The two-way tachyonic antitelephone \cite{Benford:1970xv} is the most significant thought-experiment that makes tachyons physically nonviable. To see this, consider two inertial observers $A$ and $B$ initially at point $(0, 0)$ with $A$ having a worldline $(t, vt)$ in $B$'s frame where $v < 1$. At $t = T$, $B$ decides to communicate with $A$ via a Tachyonic signal that has the worldline $(T+\lambda, u\lambda)$ in $B$'s frame where $u > 1$. In $B$'s frame, the signal must reach $A$ at $\lambda_0$ given by
\begin{align}
    &(T+\lambda_0, u\lambda_0) = (t_0, vt_0)\\
    &\implies t_0 = \frac{Tu}{u-v} \quad \lambda_0 = \frac{Tv}{u-v} 
\end{align}
Now in $A$'s frame, the worldline for the Tachyonic signal is given by a Lorentz transformation i.e.
\begin{align}
    \gamma(T+(1 - uv)\lambda, -vT+(u-v)\lambda)\quad \gamma = \frac{1}{\sqrt{1-v^2}} \label{BIT}
\end{align}
where one can have $(1-uv) < 0$ for a sufficiently high $u$. This means that in the range $(0, \lambda_0)$, $A$ will experience the signal going backwards in-time. This is a violation of causality and therefore, is a well-known paradox, called the Tolmann paradox, in tachyonic communication.

\subsection{Constraints from momentum conservation}
In order to resolve the paradox, any tachyon process must involve only tachyons that move forwards in-time. Therefore, we must consider only those processes that emit or absorb tachyons that move forward in-time in the frame of the emitter or receiver, i.e. a particle that absorbs tachyon, respectively. We parametrize the four-momenta of a tachyonic beam in $(-, +, +, +)$ signature as
\begin{align}
    &p^{\mu} =\begin{cases} 
    &\left(\frac{m}{\sqrt{u^2-1}}, \frac{m\vec{u}}{\sqrt{u^2-1}}\right) \equiv m(\Gamma, \vec{u}\Gamma)\quad \text{Forwards in-time}\\
    &\left(-\frac{m}{\sqrt{u^2-1}}, \frac{m\vec{u}}{\sqrt{u^2-1}}\right) \equiv m(-\Gamma, \vec{u}\Gamma)~ \text{Backwards in-time}
    \end{cases} \quad u > 1\notag\\
&\implies p^2 = m^2\label{ftlb}
\end{align}
where we call $m$ the mass parameter of the tachyon. In the above, we have utilized the Feynman-Stuckelberg interpretation to identify the tachyon four-momenta with negative energy as the one travelling backwards in-time \cite{thomson2013modern}. According to the Feynmann-Stuckelberg interpretation, particles that travel backwards in-time appear to us with their energies and charges having their signs flipped. That is why negative energy electrons, when traveling backwards in-time, appear to us as positrons with positive energy. In the same manner, positive energy tachyons traveling backwards in-time will appear to us as negative energy tachyons. Hence, the above parametrization. Consider now a process that involves tachyon absorption. The tachyon receiver is made up of rest mass $M$. Due to conservation of momentum, we have
 \begin{align}
      P^{\mu} + p^{\mu} = \bar{P}^{\mu} \label{AMC}
 \end{align}
where $P^{\mu}, \bar{P}^{\mu}$ are the initial and the final four-momenta of the Tachyon receiver and $p^{\mu}$ is the four-momenta of the Tachyon. We choose the following ansatz for $P^{\mu}, \bar{P}^{\mu}$
\begin{align}
    P^{\mu} = (M, 0, 0, 0)\quad \bar{P}^{\mu} = M(\gamma, \vec{w}\gamma) \quad \gamma = \frac{1}{\sqrt{1-w^2}}
\end{align}
which leads to the following equations for the different kinds of Tachyonic signals
\begin{align}
    &M(\gamma-1)-m\Gamma = 0\quad 
    m\vec{u}\Gamma = M\vec{w}\gamma \quad \text{Forwards in-time}\notag\\
    & M(\gamma-1)+m\Gamma = 0 \quad m\vec{u}\Gamma = M\vec{w}\gamma \quad\text{Backwards in-time}   \label{MCR}
\end{align}
Notice that the momentum conservation do not hold simultaneously for tachyonic signals moving forwards and backwards in-time. Therefore, for tachyons moving forward in-time, we must have $m > 0$. But for tachyonic signals of $m > 0$ moving backwards in-time, momentum conservation doesn't hold unless the receiver is made of negative mass i.e. $M < 0$. 
This implies that, in general, we must have the following selection-rule 
\begin{align}
    \operatorname{sgn}(M/m) = 1\label{C1}
\end{align}
to ensure that only the tachyonic signals that move forward in-time are absorbed in the frame of the receiver in an absorption process. Let us now discuss tachyonic emission. Consider a Tachyon emitter made of ordinary matter of rest mass $M$ emitting a tachyon. From momentum conservation, we then have
\begin{align}
    P^{\mu} = p^{\mu} + \bar{P}^{\mu} \label{Emc}
\end{align}
which leads to the following equations for the different kinds of Tachyonic signals
\begin{align}
    &M(\gamma-1)+m\Gamma = 0\quad 
    m\vec{u}\Gamma + M\vec{w}\gamma = 0 \quad \text{Forwards in time}\notag\\
    & M(\gamma-1)-m\Gamma = 0 \quad m\vec{u}\Gamma + M\vec{w}\gamma = 0 \quad\text{Backwards in time}   \label{MCE}
\end{align}
Again the momentum conservation do not hold simultaneously for tachyonic signals moving forwards and backwards in-time. 
Hence, to ensure that only the tachyonic signals that are moving forwards in-time are emitted in the frame of the emitter, we must have the following selection rule
\begin{align}
    \operatorname{sgn}(M/m) = -1\label{C2}
\end{align}
Eq. \ref{C1} and Eq. \ref{C2} constitutes a selection rule for emission and absorption of tachyons moving forwards in-time. Hence, momentum conservation implies that for a tachyonic exchange of mass parameter $m$ to always happen forwards in-time in a frame where both emitter and receiver are at rest with respect to each other, the emitter must have mass $-M$ while the receiver for the same must have mass $M$. One can also conclude that given a tachyon signal of mass $-m$, the emitter must be of mass $M$ while the receiver of the same must be of mass $-M$. But we do not consider this possibility to maintain consistency with the observation that positive masses do not emit any tachyons. 
Now, to ensure a paradox-free tachyon communication, the selection rules derived above must hold in all inertial frames of references i.e. they must be immune to the RIP. Naively, one may think that some combined representation of the tachyons, receiver and emitters can accomplish that. For instance, one can consider the following doublet representation
\begin{align}
    D= \begin{pmatrix}P_{+}\\ p\end{pmatrix} \quad E= \begin{pmatrix}P_{-}\\ -p\end{pmatrix}
\end{align}
where $D$ is the detector doublet while $E$ is the emitter doublet. $P_{\pm}$ is the four-momenta of particle with positive $(+)$ or negative $(-)$ mass at rest and $p$ is the four-momenta of the tachyon. Then, we can find a representation of the Lorentz transformation $R(\Lambda)$ such that
\begin{align}
    &R(\Lambda) D = E' \quad R(\Lambda) E = D'\notag\\
    &E' \equiv E(P'_{-}, -\Lambda p) \quad D' \equiv D(P'_{+}, \Lambda p) \quad P'_{\mp} \equiv \Lambda P_{\pm}
\end{align}
This is the RIP which we wish to avoid. We may attempt to do that by restricting to the proper orthochronous subgroup of the Lorentz group, since, such kind of Lorentz transformations change the sign of energy. However, this is still not sufficient to ensure that the sign of the tachyon energy doesn't change, see Appendix \ref{EVA}, which is the main reason for invoking the RIP. Since, the RIP cannot be avoided by representation theory, one needs to replace it with another principle in order to facilitate macroscopic tachyon mechanics. \section{Regularization principle}
Consider a generic Lorentz transformation $\Lambda$ of the form
\begin{align}
    \Lambda = 
    \bar{\gamma} \begin{pmatrix}
        1 & -\vec{v}\\
        -\vec{v} & 1
    \end{pmatrix}
\end{align}
for a tachyon beam $p^{\mu}$ moving forwards in-time with mass parameter $m > 0$ given by Eq. \ref{ftlb}
\begin{align}
    \Lambda p^{\mu} =  m(\Gamma\bar{\gamma}(1-\vec{u}\cdot\vec{v}), \Gamma\bar{\gamma}(\vec{u}-\vec{v}))
\end{align}
Now, for a frame where $(1-\vec{u}\cdot\vec{v}) < 0$, the observer sees a tachyon beam traveling backwards in-time. However, notice that we can write the above as the following infinite sum 
\begin{align}
   & m(\Gamma\bar{\gamma}(1-\vec{u}\cdot\vec{v}), \Gamma\bar{\gamma}(\vec{u}-\vec{v})) \notag\\&= \underbrace{\sum_{n=1}^{\infty}\frac{m_n}{\sqrt{\big|\frac{\vec{u}+\vec{v}_n}{1+\vec{u}\cdot\vec{v}_n}\big|^2-1}}\left(1, \frac{\vec{u}+\vec{v}_n}{1+\vec{u}\cdot\vec{v}_n}\right)}_{\text{Tail}} + \underbrace{\vphantom{\frac{m_n}{\sqrt{\big|\frac{\vec{u}+\vec{v}_n}{1+\vec{u}\cdot\vec{v}_n}\big|^2-1}}}\frac{3m\bar{\gamma}}{2}(\Gamma, \vec{u}\Gamma)}_{\text{Head}}\notag\\
    &\vec{v}_n = \frac{\vec{v}}{n^{s_0}} \quad \zeta(s_0) = -1 \quad m_n = m\frac{\bar{\gamma}}{\bar{\gamma}_n} \quad \bar{\gamma}_n = \frac{1}{\sqrt{1-|\vec{v}_n|^2}}\label{zsum1}
\end{align}
where $\zeta(s)$ is the $\zeta$-function. The LHS is obtained by $\zeta$-function regulation of the infinite sum on the RHS, see Appendix \ref{EVA}. This is analogous to the technique used in regulating the divergences in the Casimir effect \cite{Tong:2009np}. In the current context, we have used it to `regulate' a tachyon that is moving backwards in-time. In both contexts, $\zeta$-function regularization is used to assign physically meaningful interpretation to unphysical quantities. Here the unphysical quantity being the momentum of the tachyon that moves backwards in-time. Notice that each term of the series is a tachyon that is traveling forwards in-time. The part that is the Head is the fastest with mass parameter $\frac{3m\bar{\gamma}}{2}$ followed by a Tail of slower tachyons each of mass parameter $m_n$ with their speeds decreasing in the series as $n \to 1$ which is the slowest tachyon. The observer in his frame now sees the tachyon moving backwards in-time as a shower of infinite tachyons all moving forwards in-time. This ensures that emitters and receivers are not re-interpreted in any frames of reference and the paradox is resolved as $A$ in his frame will now receive a shower of tachyon particles that are moving forwards in-time.\\

 Moreover, this setup is designed in such a way that an emitter-receiver pair of mass $(-M, M)$ can only exchange a tachyon of mass parameter $m$ in their respective frames i.e. this pair is tuned for $m$ and not for arbitrary tachyon mass parameters. It is reasonable to assume that emission and absorption of tachyon particles will happen for only fixed tachyon mass parameters akin to emission and absorption spectra of photons. Therefore, in $A$'s frame, only tachyon of mass parameter $m$, which corresponds to $n=1$ in the above, will undergo detection while the rest will pass through undetected. Hence, we have acheived paradox-free tachyon communication provided we are working with a single-spectrum communication protocol. For faster-than-light multi-spectra or broadband communication protocol, this has an important consequence i.e. information encoded in tachyons undergo frame-dependent ISI or Inter-Symbol Interference, see Appendix \ref{tachISI}.

\section{Naked singularities as tachyon source}\label{NSTS}

In this section, we will see how due to this macroscopic theory, a particular class of naked singularities can emit tachyon radiation. Consider the following metric ansatz
\begin{align}
    d s^2=- e^{\nu(r)} d t^2+e^{a(r)} d r^2+r^2\left(d \theta^2+\sin ^2 \theta d \phi^2\right)
\end{align}
with the following stress-energy tensor for a tachyon
\begin{align}
    T_{\mu \nu}=-\rho(r) g_{\mu\nu} + [P(r)+\rho(r)]U_{\mu}U_{\nu} +\pi_{\mu\nu}
\end{align}
where 
\begin{align}
    &U_{\mu} = e^{a(r)/2}\delta^r_{\mu} \\ &\pi_{\mu\nu} = \operatorname{diag}(0, 0, [P(r)+\rho(r)]r^2, [P(r)+\rho(r)]r^2\sin^2\theta)
\end{align}
which satisfies 
\begin{align}
\nabla_{\mu}\pi^{\mu\nu} = 0 \quad U^{\mu}\pi_{\mu\nu} = 0 \quad U^{\mu}\nabla_{\mu}U^{\nu} = 0 \quad U^2 = 1
\end{align}
So indeed $U^{\mu}$ satisfies the geodesic equation for a tachyon. The simplest solution to the Einstein's equation 
\begin{align}
    G_{\mu\nu} = 8\pi T_{\mu\nu} 
\end{align}
is given by
\begin{align}
    ds^2 = -\left(1+\frac{2M}{r}\right)dt^2 + \frac{dr^2}{\left(1+\frac{2M}{r}\right)}+r^2\left(d \theta^2+\sin ^2 \theta d \phi^2\right)
\end{align}
which is just Schwarzschild metric with negative mass which is a well-known naked singularity \cite{Gibbons:2004au, Gleiser:2006yz, Dotti:2008ta, Belletete:2013nqa}. The above also leads to
\begin{align}
    P = 0 \quad \rho(r) = -\frac{M}{4\pi r^2}\delta(r)
\end{align} 
The divergence of the stress-energy tensor then reads
\begin{align}
    \nabla_{\mu}T^{\mu\nu} = \frac{U^{\nu}}{\sqrt{g}}\left(\partial_{r}[\sqrt{g}U^r(P+\rho)] - \sqrt{g}U^r\partial_r\rho\right) = -\delta^{\nu}_r\frac{3M^2}{4\pi r^4}\delta(r)
\end{align}
Since, the stress-energy tensor is not conserved at the centre, this can be interpreted as spontaneous creation of particles. This was the same interpretation used by Fred Hoyle \cite{hoyle1948new} to formulate his steady-state cosmological model. However, in this case, the particles being spontaneously created are tachyons. This proves that naked singularities are natural tachyon sources.  Due to Einstein's equation, it also implies that
\begin{align}
    \nabla_{\mu}G^{\mu\nu} = -\delta^{\nu}_r\frac{6M^2}{ r^4}\delta(r) = -U^{\nu}\frac{3\sqrt{2}M^\frac{3}{2}}{r^\frac{7}{2}}\delta(r)
\end{align}
which implies a violation of the Bianchi identities at the point where the naked singularity resides. This is not very surprising as such a violation also exists in the Schwarzschild metric. But there it is censored via an event horizon.
\subsection{Tachyon emission-induced censorship}

Consider a sphere of dust of negative energy density under collapse. Gravitationally, this seems difficult as negative mass particles repel each other. Even if they were to collapse in some way, the dust particles on the edge of the sphere can at random emit a tachyon away from the sphere. Notice from Eq. (\ref{MCE}), a negative mass recoils in the direction of the tachyon velocity. Therefore, the particle can exit the dust sphere and will be repelled away from the dust sphere. So, heuristically, due to tachyon emission, a sphere of dust of negative mass, even if under collapse, will lose material which can inhibit the formation of naked singularities. We will see this more rigorously. Since, we have argued that under tachyon emission a negative mass dust will lose material, therefore, the metric outside the collapsing dust would be given by the Vaidya metric as the mass of the dust sphere is changing with time. While inside the dust sphere, we will use the FLRW metric. Hence, the following metric configuration
\begin{align}
    ds^2_{+} &= -\left(1+\frac{2m(u)}{r}\right)du^2-2dudr + r^2d\Omega^2\notag\\
    ds^2_{-} &= -d\tau^2 +a^2(\tau)(d\chi^2+\sinh^2\chi d\Omega^2)\label{config1}
\end{align}
where $-$ represents the metric inside the dust of density $-\rho(\tau)$ and pressure $-P(\tau)$. $\rho(\tau)$ is defined as
\begin{align}
    \rho(\tau) \equiv \mu\frac{N(\tau)}{\frac{4}{3}\pi a^3}
\end{align}
where $N(\tau)$ is the number of particles in the dust sphere at any given time $\tau$ and is any decay function such that
\begin{align}
\lim_{\tau \to \infty}N(\tau) \to 1
\end{align}
and $-\mu$ is the mass of each dust particle. Let the coordinates at the interface $\Sigma$ be $y^{a} = (\tau, \theta, \phi)$. The metric on $\Sigma$ from outside is given by
\begin{align}
   &ds^2_{\Sigma} = -(F\dot{U}^2+2\dot{U}\dot{R})d\tau^2+ R^2d\Omega^2 \\ &F = 1+\frac{2m(u)}{R} \quad r = R(\tau) \quad u = U(\tau)
\end{align}
while the metric on $\Sigma$ within inside is given by
\begin{align}
    ds^2_{\Sigma} = -d\tau^2 + a^2(\tau)\sinh^2\chi_0d\Omega^2
\end{align}
Hence, from the above we have
\begin{align}
    F\dot{U}^2+2\dot{U}\dot{R} -1 = 0 \quad R = a(\tau)\sinh\chi_0
\end{align}
Following \cite{poisson2004relativist}, we define
\begin{align}
    e^{\mu}_a = \frac{\partial x^{\mu}}{\partial y^a}
\end{align}
Now a normal to $\Sigma$ should satisfy $n_{\mu}e^{\mu}_a = 0$. Therefore, we have
\begin{align}
   &e^{\mu+}_{0} = (\dot{U}, \dot{R}, 0, 0) \quad e^{\mu-}_{0} = (1, 0, 0, 0) \\
    &n^{+}_{\mu} = (-\dot{R}, \dot{U}, 0, 0) \quad n^{-}_{\mu} = (0, a(\tau), 0, 0) \quad n^{\pm2} = 1
\end{align}
For completeness, we compute the extrinsic curvature on both the sides of $\Sigma$ 
\begin{align}
    &K^{+}_{\tau\tau} = -\frac{1}{\dot{R}}\left[\dot{\beta}+\frac{\dot{m}}{R(\beta+\dot{R})}\right] \quad K^{+}_{\theta\theta} = \beta R \quad K^{+}_{\phi\phi} = \beta R \sin^2\theta \notag\\ &\beta = \sqrt{\dot{R}^2+F}\\
    &K^{-}_{\tau\tau} = 0 \quad K^{-}_{\theta\theta} = R\cosh\chi_0 \quad  K^{-}_{\phi\phi} =  R\cosh\chi_0\sin^2\theta
\end{align}
to compute the components of the stress-energy tensor on $\Sigma$
\begin{align}
    &[K^{\tau}_{\tau}] = \frac{1}{\dot{R}}\left[\dot{\beta}+\frac{\dot{m}}{R(\beta+\dot{R})}\right] \notag\\  &[K^{\theta}_{\theta}] = [K^{\phi}_{\phi}] =\frac{\beta-\cosh\chi_0}{R} = \frac{m(u)-\mu N(\tau)}{R^2(\beta+\cosh\chi_0)}
\end{align}
where in the above we have made use of the Einstein's equation
\begin{align}
    &a(\dot{a}^2 - 1) = -2\mu N(\tau)\\
    &\dot{a}^2+2a\ddot{a}-1 = 8\pi P(\tau)a^2
\end{align}
Since, only the particles that remain within the sphere contribute to the mass $m(u)$ at any given time, therefore, we must have
\begin{align}
    &m(u) -\mu N(\tau) = 0\\
    &\implies [K^{\theta}_{\theta}] = [K^{\phi}_{\phi}] = 0
\end{align}
which implies that the interface $\Sigma$ has some nontrivial stresses and no energy density which is consistent with a configuration that is leaking mass. However, the above system of equations has no solution corresponding to a collapsing configuration, see Appendix \ref{Einstein}. In fact, in the appendix we have shown that this method of censorship is so stringent that it doesn't allow formation of naked singularities even by artificial means. Hence, a singularity cannot form in the above metric configuration. This is generally avoided using the Weak Cosmic Censorship \cite{Tong:20019np} by invoking the dominant energy conditions in its statement which is violated by naked singularities of this type \cite{Belletete:2013nqa}. Due to the use of the Vaidya metric, it must be understood that the negative mass particles are approximated to leave the dust sphere at the speed of light. This assumption must be consistent with the kinematics of Eq. (\ref{MCE}). For a tachyon of infinite coordinate speed, we see that the recoil speed of the particle is given by
\begin{align}
    |\vec{w}| = \sqrt{\frac{m^2}{M^2+m^2}}
\end{align}
For $|\vec{w}| \approx 1$, we must have $|M| <<< m$. The above result is actually consistent with the fact that the Higgs mass which is proportional to the tachyon mass parameter is much more massive than any stable particle within the standard model i.e $m_h = \sqrt{2}\mu$, $m_h >>> m_e, m_p, m_{\nu}$ \cite{peskin2018introduction}, assuming, of course, that any stable negative  mass particles are just counterparts of stable positive mass particles within the standard model. 
Since, the tachyons that are emitted are of infinite coordinate speed, they clear the vicinity of the collapsing dust instantly. So, they will not cause any back-reactions in the configuration that may potentially contribute to the above computations and change its outcome unfavourably. This makes the result of this computation independent of the underlying field theory of the tachyons or the microscopic description of the tachyon emissions and are therefore, universal. This also implies that a tachyonic mechanism is essential for the above computations to hold.

\section{Discussion}
$\zeta$-function regularization is usually employed to regulate the unphysical divergences of QFT observables. But in this exercise, we showed that such a regularization scheme is equally effective in making sense of unphysical classical observables such as the momenta of tachyons that move backwards in-time. Notice, however, that the momenta of tachyon moving backwards in-time is a finite quantity while its regularized version is an infinite divergent sum of tachyon momenta that moves forwards in-time i.e. an unphysical finite quantity is regulated as a physical divergence in this scheme. This is opposite to how the regularization scheme works in QFT.\\

We also had to employ the usage of negative mass in order to formulate a consistent tachyon mechanics which may be considered somewhat unphysical from the particle physics perspective. However, one can run the argument backwards where one can perform the GR computation in Section \ref{NSTS} first and acknowledge the issue of naked singularities being potential tachyon sources. As there are a priori no principles or arguments to disregard such computations in GR as unphysical, this can lead to all sorts of issues in classical GR with regards to global hyperbolicity. Hence, the need for a more stringent censorship principle arises that doesn't allows formation of such singularities. In the literature as well, negative masses are generally considered as exotic matter \cite{bondi1957negative} rather than unphysical as they are demonstrated to have physical applications in dark matter, dark energy \cite{petit2014negative} and cosomology \cite{Manfredi:2018nlx}. From this perspective, the usage of negative masses as part of a consistent tachyonic framework becomes more acceptable.\\

\section{Summary}
In summary, we discussed the means by which we can have a consistent macroscopic tachyon mechanics where the temporal paradoxes are resolved. By first imposing momentum conservation, we derived the selection rules for emitters and receivers that emit and absorb, respectively, only tachyons that move forwards in-time in their respective frames. In order to have a viable tachyon mechanics, these selection rules must hold in all frames which implied that emitters and receivers must be immune to reinterpretation. We showed that this cannot be guaranteed by representation theory alone but only by postulating a regularization principle of tachyons that move backwards in-time as a tachyon shower that move forwards in-time. We achieved this by representing the tachyon moving backwards in-time as an infinite sum of tachyons that move forwards in-time using the properties of the $\zeta$-function. This effectively resolves all temporal paradoxes as tachyons move forwards in-time in all inertial frames either as a single particle or a particle shower. Later we showed that such a mechanics, if true, allows for certain naked singularities to be natural sources of tachyon radiation. However, it turned out that the same mechanics did not allow gravitational collapse solutions in GR that can lead to the formation of such naked singularities. Thus also demonstrating the role of paradox-free tachyons in cosmic censorship.
\section{Acknowledgements}
I would like to acknowledge the support and kindness of my supervisor and employer Prof. Junggi Yoon (JRG, Holography and Black Holes), APCTP, Pohang, Republic of Korea without which this initiative would not have been possible. APCTP is supported through the Science and Technology Promotion Fund and Lottery Fund of the Korean Government. This research is dedicated to the people of India and the Republic of Korea for their steady support of research in theoretical science. This work was supported by the NRF grant funded by the Korea government (MSIT) (No. 2022R1A2C1003182).

\section*{Declarations}

\begin{itemize}
\item Funding:  This work was supported by the NRF grant funded by the Korea government (MSIT) (No. 2022R1A2C1003182).
\item Conflict of interest/Competing interests (check journal-specific guidelines for which heading to use): Not applicable
\item Ethics approval and consent to participate: Yes
\item Consent for publication: Yes
\item Data availability: Not applicable 
\item Materials availability: Not applicable
\item Code availability: Not applicable 
\item Author contribution: Full
\end{itemize}

\begin{appendices}

\section*{Appendix}
\section{Solution to the Einstein's equations}\label{Einstein}
Consider the Einstein's equations for the following metric
\begin{align}
    ds^2 &= -d\tau^2 +a^2(\tau)(d\chi^2+\sinh^2\chi d\Omega^2)
\end{align}
with the following stress-energy tensor
\begin{align}
    T^{\mu}_{~\nu} = \operatorname{diag}(-\rho, -P, -P, -P)
\end{align}
which are given by
\begin{align}
    &a(\dot{a}^2 - 1) = -2\mu N(\tau) \label{E1}\\
    &\dot{a}^2+2a\ddot{a}-1 = 8\pi P(\tau)a^2\label{E2}
\end{align}
which can be rearranged to give
\begin{align}
    \mu\dot{N} = -4\pi a^2\dot{a} P
\end{align}
Now, the above be solved to give
\begin{align}
    \frac{4}{3}\pi a^3 = C -\mu\int^N \frac{dN'}{P(N')} \quad C > 0
\end{align}
Assuming that negative mass particles are only exotic in the sense that they have negative mass and are not unusual in any other sense, then the equation of state must follow $P(N) \to 0$ as $N \to 1$. Let the leading behaviour as of $P(N)$ be given by
\begin{align}
    P(N) = \epsilon (N-1)^{\alpha} \quad \alpha > 0
\end{align}
as $N \to 1$. Then as $N \to 1$, we have
\begin{align}
     V \equiv \frac{4}{3}\pi a^3 = C -\mu\frac{(N-1)^{1-\alpha}}{(1-\alpha)\epsilon}
\end{align}
However, notice that $V$ increases as $N$ decreases which means that the dust sphere becomes larger as it loses the dust particles due to tachyon emission implying that this set of Einstein's equations do not have a collapsing solution. Contrast this with the case where there is no tachyonic emission one can `artificially' force the negative mass particles to collapse into a naked singularity. To see this consider
\begin{align}
    &(\dot{a}^2 - 1) = -\frac{8\pi}{3}\rho(\tau) a^2 \equiv -2\frac{\mu(\tau)}{a} \label{E1}\\
    &\dot{a}^2+2a\ddot{a}-1 = 8\pi P(\tau)a^2\label{E2}
\end{align}
which can be rearranged to give
\begin{align}
   \dot{\mu} = -4\pi P a^2\dot{a}
\end{align}
which when solved leads to
\begin{align}
    \frac{4}{3}\pi a^3 = C-\int^{\mu} \frac{d\mu'}{P(\mu')} \quad C > 0
\end{align}
If
\begin{align}
P(\mu) = \epsilon\mu^{\beta} \quad 0 < \beta < 1
\end{align}
then we have
\begin{align}
    \frac{4}{3}\pi a^3 &= C-\frac{\mu^{1-\beta}}{\epsilon(1-\beta)}\\
    \implies \rho &= \frac{\mu}{\frac{4}{3}\pi a^3} = \frac{\epsilon(1-\beta)\mu}{\epsilon(1-\beta)C-\mu^{1-\beta}}
\end{align}
Since, $\mu(\tau)$ is an increasing function of $\tau$, at some $\tau = \tau_0$, we will have $\rho \to\infty$ which is a naked singularity. This can be made possible via the following metric configuration
\begin{align}
    ds^2_{+} &= -\left(1+\frac{2m(v)}{r}\right)dv^2+2dvdr + r^2d\Omega^2\notag\\
    ds^2_{-} &= -d\tau^2 +a^2(\tau)(d\chi^2+\sinh^2\chi d\Omega^2)\label{config_art}
\end{align}
However, if tachyonic emissions by negative mass exists, then $\mu(\tau)$ can no longer be guaranteed to be an increasing function of $\tau$. Apart from this, there are no arguments or processes within GR itself that can prevent this. Therefore, tachyonic emission enforces censorship in natural GR processes by forcing $N(\tau)$ to be a decreasing function in $\tau$ with the metric in Eq. (\ref{config1}). If one tries to bypass this natural censorship by some 'artificial' means using Eq. (\ref{config_art}) to force a naked singularity, the tachyon emission again enforces censorship by guaranteeing that $\mu(\tau)$ is not an increasing function in $\tau$.
\section{Regularization of momenta for tachyon traveling backwards in-time}\label{EVA}
Consider the momentum conservation for tachyon absorption given by Eq. \ref{AMC} under a Lorentz transformation $\Lambda$
\begin{align}
     & \Lambda P^{\mu} + \Lambda p^{\mu} = \Lambda \bar{P}^{\mu}\notag\\
    & \implies M(\bar{\gamma}, -\vec{v}\bar{\gamma}) + m(\Gamma\bar{\gamma}(1-\vec{u}\cdot\vec{v}), \Gamma\bar{\gamma}(\vec{u}-\vec{v})) \notag\\&~~~~~~~~~= M(\gamma\bar{\gamma}(1-\vec{v}\cdot\vec{w}), \gamma\bar{\gamma}(\vec{w}-\vec{v})) \label{TAR}\\ 
    &\text{where}~~ \Lambda = 
    \bar{\gamma} \begin{pmatrix}
        1 & -\vec{v}\\
        -\vec{v} & 1
    \end{pmatrix}
    \notag
\end{align}
Notice that when $\vec{u}\parallel \vec{v}$
\begin{align}
    &\gamma\bar{\gamma} (1-\vec{v}\cdot\vec{w}) = \frac{1}{\sqrt{1-\big |\frac{\vec{w}-\vec{v}}{1-\vec{v}\cdot\vec{w}}}\big|^2} \quad \text{for} ~~\vec{v} \parallel \vec{w}\\
    &\Gamma\bar{\gamma}(1-\vec{u}\cdot\vec{v}) = \begin{cases} 
    &\frac{1}{\sqrt{\big |\frac{\vec{u}-\vec{v}}{1-\vec{u}\cdot\vec{v}}\big|^2-1}}\quad (1 -\vec{u}\cdot\vec{v}) > 0\\
    &-\frac{1}{\sqrt{\big |\frac{\vec{u}-\vec{v}}{\vec{u}\cdot\vec{v}-1}\big|^2-1}}\quad (1 -\vec{u}\cdot\vec{v}) < 0
    \end{cases}
\end{align}
When $(1 -\vec{u}\cdot\vec{v}) > 0$, we simply have a tachyon absorption process involving initially a positive mass $M$ with speed $|\vec{v}|$ in a head-on collision with a tachyon of speed $\big |\frac{\vec{u}-\vec{v}}{1-\vec{u}\cdot\vec{v}}\big|$ which after absorption leads to the receiver particle acquire a speed of $\big |\frac{\vec{w}-\vec{v}}{1-\vec{v}\cdot\vec{w}}\big|$. In the case of $(1 -\vec{u}\cdot\vec{v}) < 0$, we now have a tachyon that is traveling backwards in-time. However, let us split the second term in the LHS as follows
\begin{align}
   & m(\Gamma\bar{\gamma}(1-\vec{u}\cdot\vec{v}), \Gamma\bar{\gamma}(\vec{u}-\vec{v})) \notag\\&= m(\Gamma\bar{\gamma}(\zeta(0)+\zeta(s_0)\vec{u}\cdot\vec{v}), \Gamma\bar{\gamma}(\zeta(0)\vec{u}+\zeta(s_0)\vec{v})) +   \frac{3m\bar{\gamma}}{2}(\Gamma, \vec{u}\Gamma)\notag\\&= \sum_{n=1}^{\infty}m(\Gamma\bar{\gamma}(1+\vec{u}\cdot\vec{v}_n), \Gamma\bar{\gamma}(\vec{u}+\vec{v}_n)) + \frac{3m\bar{\gamma}}{2}(\Gamma, \vec{u}\Gamma) \notag\\&= \sum_{n=1}^{\infty}m(\Gamma\bar{\gamma}(1+\vec{u}\cdot\vec{v}_n), \Gamma\bar{\gamma}(\vec{u}+\vec{v}_n)) + \frac{3m\bar{\gamma}}{2}(\Gamma, \vec{u}\Gamma) = \notag\\
    & = \sum_{n=1}^{\infty}m_n(\Gamma\bar{\gamma}_n(1+\vec{u}\cdot\vec{v}_n), \Gamma\bar{\gamma}_n(\vec{u}+\vec{v}_n)) + \frac{3m\bar{\gamma}}{2}(\Gamma, \vec{u}\Gamma)\label{zsum} \\&= \underbrace{\sum_{n=1}^{\infty}\frac{m_n}{\sqrt{\big|\frac{\vec{u}+\vec{v}_n}{1+\vec{u}\cdot\vec{v}_n}\big|^2-1}}\left(1, \frac{\vec{u}+\vec{v}_n}{1+\vec{u}\cdot\vec{v}_n}\right)}_{\text{Tail}} + \underbrace{\vphantom{\frac{m_n}{\sqrt{\big|\frac{\vec{u}+\vec{v}_n}{1+\vec{u}\cdot\vec{v}_n}\big|^2-1}}}\frac{3m\bar{\gamma}}{2}(\Gamma, \vec{u}\Gamma)}_{\text{Head}}\\
    &\vec{v}_n = \frac{\vec{v}}{n^{s_0}} \quad \zeta(s_0) = -1 \quad m_n = m\frac{\bar{\gamma}}{\bar{\gamma}_n} \quad \bar{\gamma}_n = \frac{1}{\sqrt{1-|\vec{v}_n|^2}}
\end{align}
where $\zeta(s)$ is the $\zeta$-function. Each term in the infinite sum is now a tachyon moving forward in-time. This is the proposed regularization principle. The same principle works for the emission process as well.\\
\section{Inter-Symbol Interference (ISI) in tachyon signals traveling backwards in-time}\label{tachISI}
Consider two obsevers $A$ and $B$ each with a worldline $(t, vt)$ and $(t, 0)$, respectively. Now, $B$ sends a tachyon signal to $A$ which has a worldline $(T+\lambda, u\lambda)$ with respect to $B$. Now, with respect to $A$, the tachyon signal worldline looks like
\begin{align}
    S: \bar{\gamma}(T+\lambda (1-uv), -vT+ (u-v)\lambda)
\end{align}
The worldline of $B$ with respect to $A$ is given by 
\begin{align}
    B: \bar{\gamma}(t, -vt)
\end{align}
With respect to $B$, the signal reaches $A$ at
\begin{align}
    t_0 = \frac{Tu}{u-v} \quad \lambda_0 = \frac{Tv}{u-v} 
\end{align}
When $(1-uv) < 0$, $A$ will see the signal traveling backwards in-time. However, as postulated before, in $A$'s frame with $(1-uv) < 0$, the past-pointing tachyon worldline can be represented as an infinite sum of future-pointing tachyon worldlines using the $\zeta$-function regularization. These worldlines can be inferred from Eq. (\ref{zsum}) and are given by\footnote{$T-$Tail, $H-$Head} 
\begin{align}
    S^{(n)}_T: \bar{\gamma}_n\left(\frac{\bar{\gamma}}{\bar{\gamma}_n}T+\lambda (1+uv_n), -\frac{\bar{\gamma}}{\bar{\gamma}_n} vT+ (u+v_n)\lambda\right) \quad S_H: (\bar{\gamma}T+\lambda, -\bar{\gamma}vT+u\lambda)
\end{align}
such that $d\lambda/d\tau = \Gamma$. Now, in $A$'s frame, they reach $A$ at different points of time which are as follows
\begin{align}
    t'_\infty = \bar{\gamma}\left(\frac{u+v}{u}\right)T \quad t'_n = t'_{\infty}\left[1 + \frac{(u^2-1)vv_n}{(u+v)(u+v_n)}\right]
\end{align}
Hence, we cannot assign any single time at which $A$ will receive the signal. But one can assign a range in which the entirety of the signal will reach $A$ and is given by
\begin{align}
\Delta t' = t'_1-t'_{\infty} = \bar{\gamma}T\frac{(u^2-1)v^2}{u(u+v)}
\end{align}
This is very reminiscent of the pulse width in the field of signal processing. Notice that $t'_i > 0$ for all $i$'s. Therefore, in frames where $(1-uv) < 0$, tachyon signals simply gain a pulse width of $\Delta t'$ instead. 
\begin{figure}[h!]
\centering
    \includegraphics[scale=0.317]{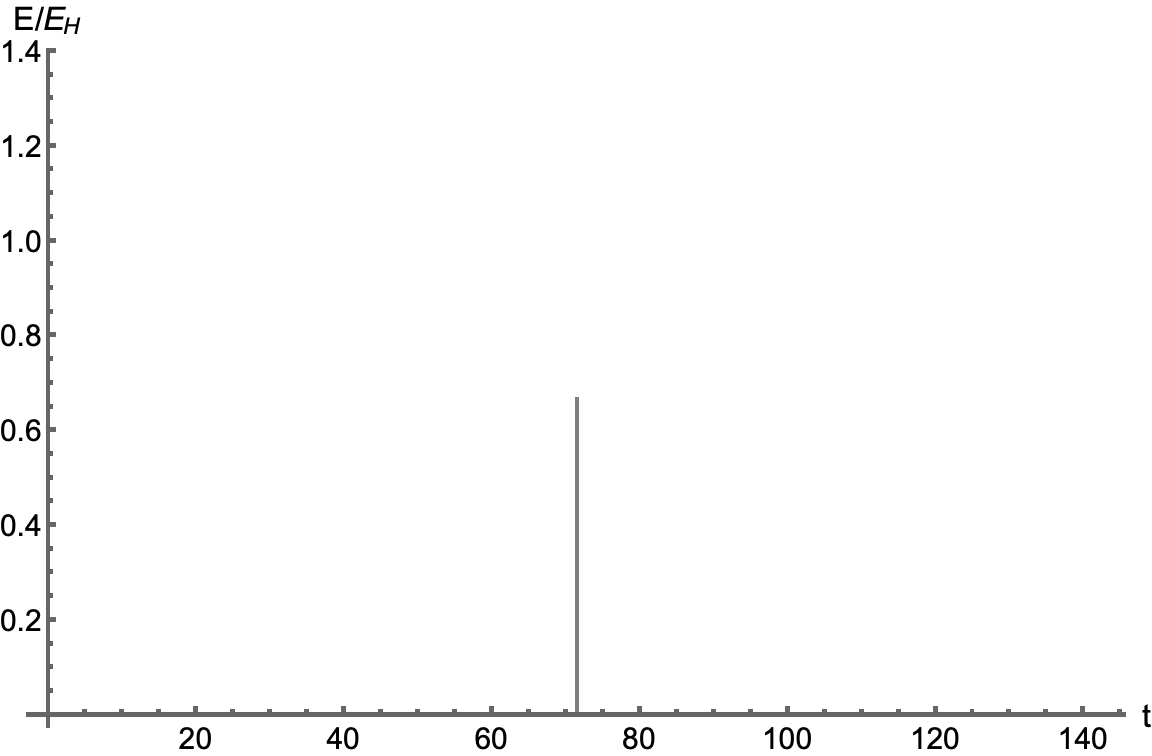}
     \includegraphics[scale=0.317]{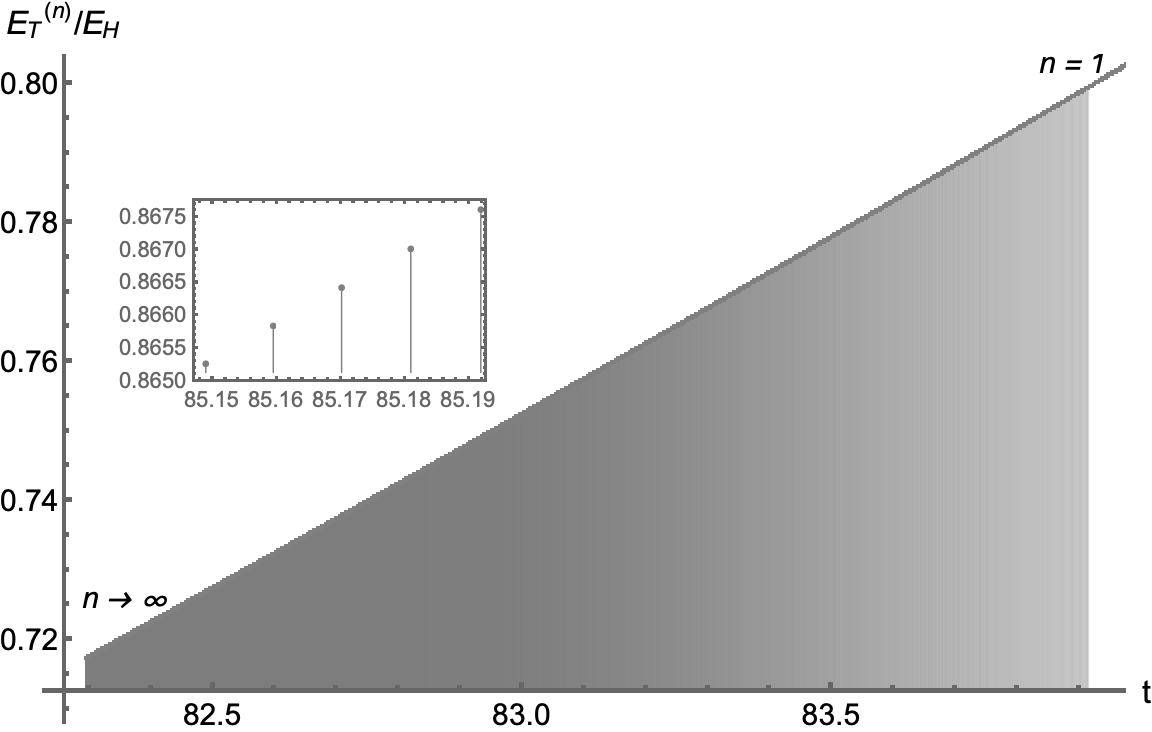}
     \caption{\emph{Left:} A single tachyon pulse of a definite energy with zero pulse width. \emph{Right:} The same pulse observed in $A$'s frame now has a nonzero pulse width. The entire pulse is made up of individual components of the \emph{tachyon shower}. The energy is normalized with respect to the energy of the Head tachyon $E_H$. The energy of the components increase as $n \to 1$.
     }\footnotemark\label{LRP}
\end{figure}\footnotetext{These graphs are generated for the following values: $u = 2, v = 0.6, s_0 \approx 0.3, T = 50$\label{vals}}
\begin{figure}[h!]
\centering
    \includegraphics[scale=0.317]{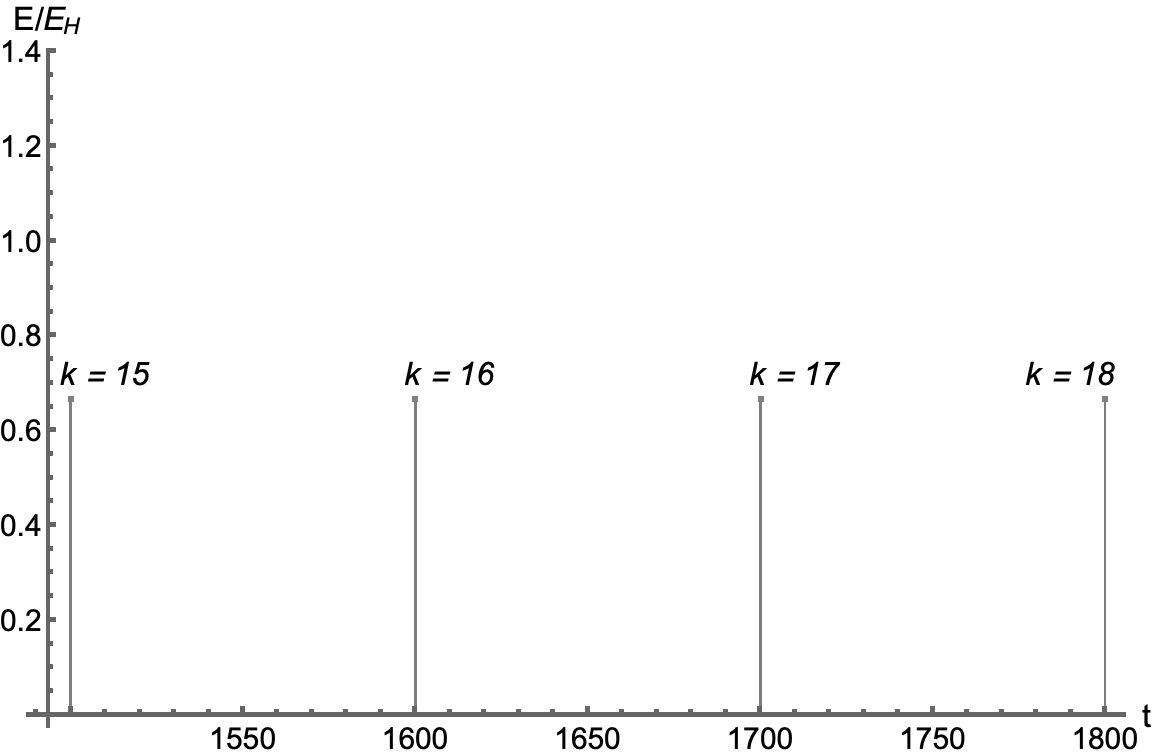}
     \includegraphics[scale=0.317]{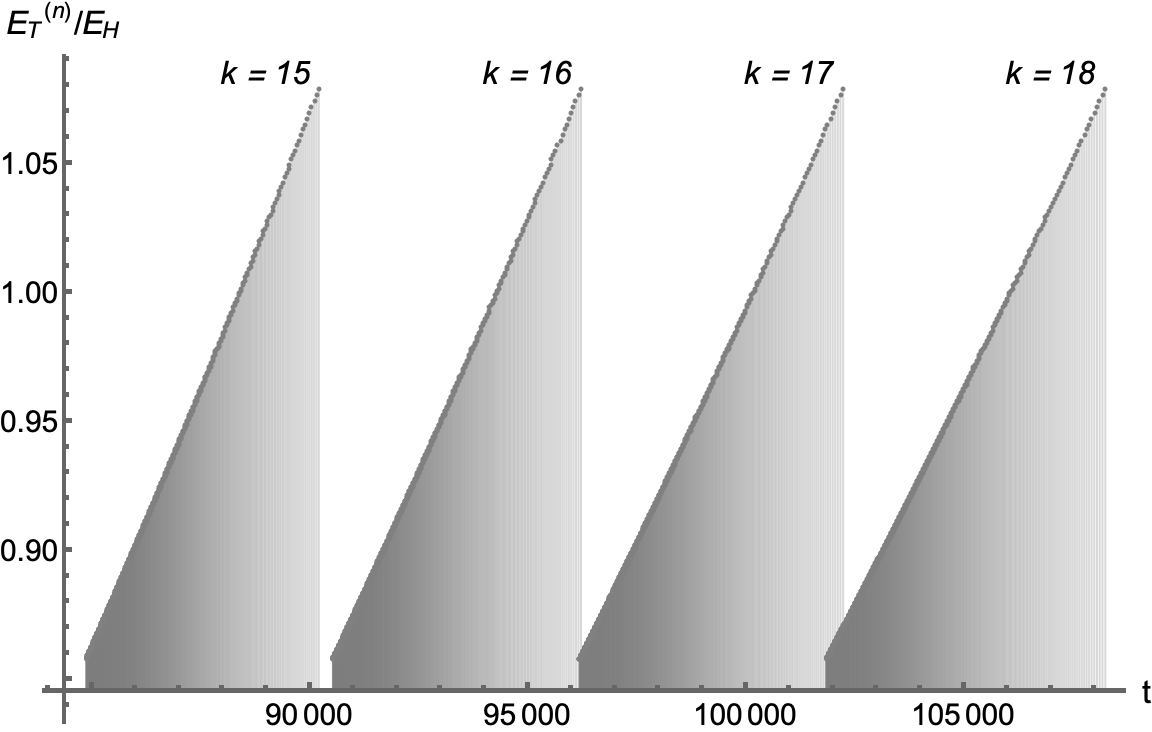}
     \caption{\emph{Left:} A pulsed tachyon signal in $B$'s frame. \emph{Right:} The same signal observed in from $A$'s frame. Along with the gain in the pulse width there is overlapping of pulses at later intervals which is an inevitable consequence of Eq. (\ref{overlapcon}). As $k$ becomes larger, the overlap becomes more significant.}\footnotemark\label{LRP2}
\end{figure} \footnotetext{These graphs are generated for the following values: $u = 2, v = 0.9999, s_0 \approx 0.3, T = 50$. In telecommunications terminology, this is called Inter-Symbol Interference (ISI). In the \emph{Left} image if you assign each pulse a $1$ and the gap between each pulse a $0$, then these are called symbols. In the \emph{Right} image, you can then see that the symbols interfere as the pulse width changes and the pulses begin to overlap.}
See Figure \ref{LRP}. Now, we consider a pulsed tachyon signals sent to $B$ from $A$ with a periodic time interval $T$. In the frame of $B$, the worldline of the $k$-th pulse looks like
\begin{align}
     S_k: \bar{\gamma}(kT+\lambda(1-uv), -kvT + (u-v)\lambda)
\end{align}
Again due to the regularization principle, we have
\begin{align}
    &S^{(n)}_{k,T}: \bar{\gamma}_n\left(\frac{\bar{\gamma}}{\bar{\gamma}_n}kT+\lambda (1+uv_n), -\frac{\bar{\gamma}}{\bar{\gamma}_n} kvT+ (u+v_n)\lambda\right) \notag\\ &S_{k,H}: (\bar{\gamma}kT+\lambda, -\bar{\gamma}kvT+u\lambda)
\end{align}
The pulse width for the $k$-th pulse is given by
\begin{align}
   &t'_{k, \infty} = k \bar{\gamma}\left(\frac{u+v}{u}\right)T \quad t'_{k, n} = t'_{k, \infty} \left[1 + \frac{(u^2-1)vv_n}{(u+v)(u+v_n)}\right]\notag \\&\Delta t'_k = t'_{k,1}-t'_{k, \infty} =\bar{\gamma}kT\frac{(u^2-1)v^2}{u(u+v)}
\end{align}
Since, the pulse has fragmented into infinite components, the periodic time interval for the $n$-th component is given by 
\begin{align}
   T^{'(n)} = t'_{1, \infty} \left[1 + \frac{(u^2-1)vv_n}{(u+v)(u+v_n)}\right]
\end{align}
Notice that
\begin{align}
    T^{'(n)} > T^{'(n+1)}\label{overlapcon}
\end{align}
Since, the fragments do not have the same time interval, this means that a periodically-pulsed tachyon signal will appear to be aperiodic and will eventually appear to overlap at later intervals to the moving observer, see Figure \ref{LRP2}. The observed changes in the pulse width along with overlapping of pulses in $A$'s frame makes any information encoded in the tachyon signals difficult to recover, especially, in broadband communications. Therefore, one may need to device novel information recovery protocols to mitigate these effects.

\end{appendices}

\printbibliography

\end{document}